\documentclass[aps,prb,preprint,showpacs,superscriptaddress,amsmath,amssymb]{revtex4}
\pdfoutput=1
\usepackage{graphicx}
\usepackage{dcolumn}
\usepackage{bm}
\usepackage{color}
\usepackage{ifpdf}
\RequirePackage[english]{babel}

\begin{document}

\title{Terahertz Transverse-Electric- and Transverse-Magnetic-polarized waves localized on graphene in photonic crystals}

\author{Yu.O.~Averkov}
\affiliation{ A.Ya.~Usikov Institute for Radiophysics and Electronics, National Academy of Sciences of Ukraine, 61085 Kharkov, Ukraine}
\email{yuriyaverkov@gmail.com}

\author{V.M.~Yakovenko}
\affiliation{ A.Ya.~Usikov Institute for Radiophysics and Electronics, National Academy of Sciences of Ukraine, 61085 Kharkov, Ukraine}

\author{V.A.~Yampol'skii}
\affiliation{ A.Ya.~Usikov Institute for Radiophysics and Electronics, National Academy of Sciences of Ukraine, 61085 Kharkov, Ukraine}
\affiliation{V.N.~Karazin Kharkov National University, 61077 Kharkov, Ukraine}
\affiliation{CEMS, RIKEN, Saitama 351-0198, Japan}

\author{Franco Nori}
\affiliation{CEMS, RIKEN, Saitama 351-0198, Japan}
\affiliation{Department of Physics, University of Michigan, Ann Arbor, Michigan 48109, USA}

\begin{abstract}
We predict the coexistence of both TE- and TM-polarized localized electromagnetic waves that can propagate \emph{in the same frequency range} along a graphene layer inserted in a photonic crystal. In addition, we studied the excitation of these modes by an external wave and have shown that the resonance peaks of the sample transmissivity should be observed due to the excitation of the localized waves, independently of the polarization of the exciting wave. The simplicity of the derived dispersion relations for the localized modes and the possibility to excite waves of both polarizations provide a method for measuring graphene conductivity.
\end{abstract}


\maketitle
\section{INTRODUCTION}

The unusual and remarkable transport properties of graphene have attracted considerable attention, including: an unconventional quantum Hall effect~\cite{HE}; the possibility of testing the Klein paradox~\cite{KP}; the Aharonov-Bohm effect in graphene rings~\cite{AB}, as well as mesoscopic effects, such as weak localization~\cite{WL}, conductance fluctuations~\cite{CF}, quantum noise~\cite{QN}, Coulomb blockade~\cite{CB}, and Anderson localization~\cite{AL}; specular Andreev reflection and Josephson effect~\cite{AR}; formation of a Wigner crystal~\cite{WC}; voltage-driven quantum oscillations of the conductance~\cite{VD}; intriguing electron lensing ~\cite{EL}, and other fascinating phenomena (see, e.g., Refs.~\onlinecite{4,9a,1,2,3,5,6,7,8,9} and references therein). Studies of graphene are also inspired by its potential application in nanoelectronic devices, because the electron concentration can vary considerably due to applied electric fields, and graphene can have both electrons and holes as high-mobility charge carriers.

A main feature of the graphene electron structure, which is very different from conventional two-dimensional electron systems, is the existence of six Dirac cones at the corners of an hexagon-shaped Brillouin zone with a massless linear electron-hole dispersion. This specific spectrum for the charge carriers leads to a number of interesting transport properties, or imparts new features to them. The phenomena listed above are caused by the quantum peculiarities of graphene, and these manifest at the \emph{quantum} level. However, it is worthwhile to emphasize that the quantum features of the graphene conductivity can also play a very important role in \emph{classical} macroscopic phenomena. A nontrivial example of a classical phenomenon is the propagation of the electromagnetic waves localized near a graphene layer inserted into a dielectric~\cite{Mikhailov}. It is known that surface electromagnetic waves cannot propagate along the interface between two dielectrics. Surprisingly, the addition of only one monatomic graphene layer changes radically the \emph{macroscopic} electrodynamic properties of the system.  The graphene layer can support a surface wave between two dielectrics and can play a very important role in other problems of plasmonics (see, e.g., Refs.~\onlinecite{Garcia,10,11,12,13,14,15,16,17}).

The Dirac spectrum of electrons leads to new features of the electrodynamic response of the electron-hole plasma in graphene, as compared to conventional electron systems. For instance, the existence of a localized transverse-electric (TE) mode, a mode which cannot exist in systems with a parabolic electron dispersion, was predicted in Ref.~\onlinecite{Mikhailov} for the graphene in a \emph{symmetrical dielectric environment}. Thus, both the transverse magnetic (TM) and transverse electric surface waves can propagate along the graphene layer. However, these TM and TE modes exist in very different frequency ranges. Indeed, as shown in Ref.~\onlinecite{Mikhailov}, the frequency range for the TE modes can vary from radio to infrared frequencies, depending on the carriers concentration, but the TM surface waves do not exist  at the frequencies for the TE modes.

In this paper,  we predict the coexistence of localized both TE- and TM-polarized electromagnetic waves that can propagate \emph{in the same frequency range} along the graphene layer inserted into a photonic crystal (PC).

It is important to emphasize that the TE-polarized localized waves can propagate along the graphene layer placed between two semi-infinite dielectrics \emph{with very similar} permittivities only. Here we consider the case when the graphene layer is in a \emph{non-symmetrical environment}. We show that, due to the periodic structure of the environment, localized TE-polarized waves can exist even in such a non-symmetric case. Moreover, this periodicity also allows the possibility for propagation of the TE- and TM-polarized localized waves in the same frequency range, contrary to the case of  propagation of the TE- and TM-polarized waves along the graphene layers placed between two identical semi-infinite dielectrics.

We also consider the problem of the  excitation of the TE- and TM-polarized modes by the external wave that irradiates the PC-graphene-PC structure. We show that the resonance peak of the transmissivity of the structure appears, when changing the frequency or the incident angle, due to the TM or TE surface waves excitation, independently of the polarization of the exciting wave. The predicted phenomenon can be observed in the terahertz frequency range,  which is very important for various applications but not easily accessible with modern electronic and optical devices. This technological perspective provides an additional motivation for study of these phenomena. The analysis of the resonance peaks of the wave transmissivity can give important information on the graphene conductivity in the centimeter, millimeter, and submillimeter wavelength frequency ranges.

\section{Dispersion relations for the TE- and TM-polarized surface waves}

\begin{figure}
\includegraphics [width=15.0 cm,height=11.0 cm]{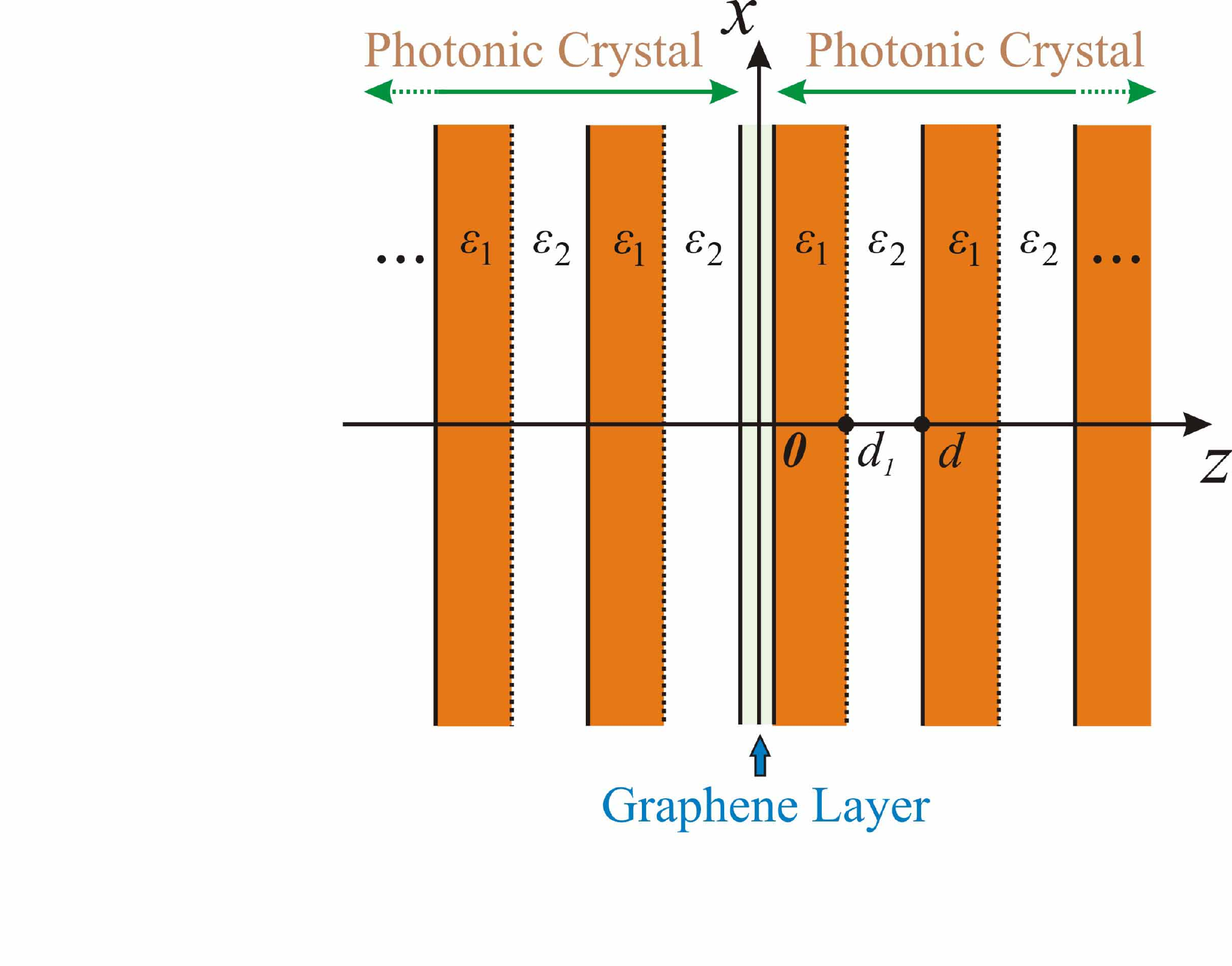}
\caption{\label{Fig1} (Color online) Geometry of the problem for waves localized on the graphene layer.}
\end{figure}

Consider a graphene layer inserted into a photonic crystal (PC). The elementary cell consists of two nonmagnetic dielectrics with permittivities $\varepsilon_1$, $\varepsilon_2$ and thicknesses $d_1$, $d_2$, respectively. The period of the PC structure is $d=d_1+d_2$. The $z$-axis is perpendicular to the layers of the PC, and the graphene layer is arranged in the plane $z=0$. Thus, the photonic crystal occupies the half-spaces $z<0$ and $z>0$ (see Fig.~\ref{Fig1}).

First we consider a TM-polarized surface wave with wave vector ${\vec k}=(k_x,0,k_z)$ and with the following components of the electric ${\vec E}$ and magnetic ${\vec H}$ fields proportional to $\exp(-i\omega t)$: ${\vec E}=(E_x,0,E_z)$ and ${\vec H}=(0,H_y,0)$. The fields in the photonic crystal satisfy the translation condition~\cite{Yeh},
\begin{equation}\label{eq1}
   \begin{pmatrix} H_y(z=(m+1)d)  \\ E_x(z=(m+1)d) \end{pmatrix}=\mathrm{\bf M}^{\rm TM} \begin{pmatrix} H_y(z=md)  \\ E_x(z=md) \end{pmatrix},\quad
\end{equation}
and the Bloch relation
\begin{equation}\label{eq2}
   \begin{pmatrix} H_y(z=(m+1)d)  \\ E_x(z=(m+1)d) \end{pmatrix}=\exp(\pm i \varphi) \begin{pmatrix} H_y(z=md)  \\ E_x(z=md) \end{pmatrix},\quad
\end{equation}
where $m$ is an integer number, $\mathrm{\bf M}^{\rm TM}$ is the propagation matrix for the TM wave with the elements,
\begin{equation}\label{eq3}
    \mathrm{M}_{11}^{\rm TM}=\cos\varphi_1 \cos\varphi_2 - \displaystyle\frac{\varepsilon_2 k_{1z}}{\varepsilon_1 k_{2z}}\sin\varphi_1 \sin\varphi_2,
\end{equation}
\begin{equation}\label{eq4}
    \mathrm{M}_{12}^{\rm TM}=i\displaystyle\frac{\omega \varepsilon_1}{c k_{1z}}\sin\varphi_1 \cos\varphi_2 + i\displaystyle\frac{\omega \varepsilon_2}{c k_{2z}}\sin\varphi_2 \cos\varphi_1,
\end{equation}
\begin{equation}\label{eq5}
    \mathrm{M}_{21}^{\rm TM}=i\displaystyle\frac{c k_{1z}}{\omega \varepsilon_1}\sin\varphi_1 \cos\varphi_2 + i\displaystyle\frac{c k_{2z}}{\omega \varepsilon_2}\sin\varphi_2 \cos\varphi_1,
\end{equation}
\begin{equation}\label{eq6}
    \mathrm{M}_{22}^{\rm TM}=\cos\varphi_1 \cos\varphi_2 - \displaystyle\frac{\varepsilon_1 k_{2z}}{\varepsilon_2 k_{1z}}\sin\varphi_1 \sin\varphi_2,
\end{equation}
where $\varphi_j=k_{jz}d_j$ ($j=1,2$), $k_{jz}=\sqrt{ k_0^2\varepsilon_j-k_x^2}$, $k_0=\omega/c$, $\omega$ is the wave frequency, $\varphi=q d$, $q$ is the complex Bloch number, and $\cos\varphi=(\mathrm{M}_{ 11}^{\rm TM}+\mathrm{M}_{22}^{\rm TM})/2$~\cite{Yeh}.
For the determinacy, we assume that $\mathrm{Im} (q) >0$. In this case, the signs ``+'' and ``$-$'' in the exponent in Eq.~\eqref{eq2} correspond to the electromagnetic fields in the regions $z>0$ and $z<0$, respectively.

The boundary conditions on the graphene layer (at $z=0$) consist of the continuity of the component $E_x$ of the electric field and the presence of the jump of the magnetic field component $H_y$, caused by the current in graphene,
\begin{equation}\label{eq9b}
     H_y(+0)-H_y(-0)=-\displaystyle\frac{4\pi \sigma}{c} E_x(0).\
\end{equation}
Here $\sigma = \sigma^{\rm intra}+\sigma^{\rm inter}$ is the graphene conductivity, which is the sum of the intraband conductivity $\sigma^{\rm intra}$ and interband conductivity $\sigma^{\rm inter}$~\cite{Falk_JETP_2008}. For a degenerate electron gas, when $k_B T \ll \mu$ (here $k_B$ is the Boltzmann constant, $T$ is the graphene temperature, and $\mu$ is the chemical potential), $\sigma^{\rm intra}$ and $\sigma^{\rm inter}$ can be written as~\cite{Falk_JETP_2008},
 \begin{equation}\label{eq10}
    \sigma^{\rm intra}=\displaystyle\frac{i e^2 \mu}{\pi \hbar^2 (\omega+i \nu)},\
 \end{equation}
 \begin{equation}\label{eq11}
    \sigma^{\rm inter}=\displaystyle\frac{e^2 }{4 \hbar} \Bigl[\theta (\hbar \omega - 2 \mu) - \displaystyle\frac{i}{2\pi}
    \ln\displaystyle\frac{(\hbar\omega+2\mu)^2}{(\hbar\omega-2\mu)^2+(2 k_B T)^2}\Bigr],\
 \end{equation}
 where $\nu$ is the intraband electron relaxation frequency and $\theta(x)$ is the Heaviside step function.

 Using Eqs.~(\ref{eq1}--\ref{eq9b}), we obtain the following dispersion relation for the TM-polarized surface wave:
 \begin{equation}\label{eq14}
    \sin\varphi=\displaystyle\frac{2\pi i\sigma}{c} \mathrm{M}_{21}^{\rm TM}.
 \end{equation}

 Similarly, we can derive the dispersion relation for the TE-polarized surface wave with the following components of the electromagnetic field and the wave vector:  ${\vec E}=(0,E_y,0)$, ${\vec H}=(H_x,0,H_z)$, and ${\vec k}=(k_x,0,k_z)$.
 The fields in the photonic crystal satisfy the conditions~\cite{Yeh},
 \begin{equation}\label{eq15}
   \begin{pmatrix} E_y(z=(m+1)d)  \\ H_x(z=(m+1)d) \end{pmatrix}=\mathrm{\bf M}^{\rm TE} \begin{pmatrix} E_y(z=md)  \\ H_x(z=md) \end{pmatrix},\quad
\end{equation}
\begin{equation}\label{eq16}
   \begin{pmatrix} E_y(z=(m+1)d)  \\ H_x(z=(m+1)d) \end{pmatrix}=\exp(\pm i \varphi) \begin{pmatrix} E_y(z=md)  \\ H_x(z=md) \end{pmatrix} \quad
\end{equation}
where $\mathrm{\bf M}^{\rm TE}$ is the propagation matrix for the TE wave with components:
\begin{equation}\label{eq17}
    \mathrm{M}_{11}^{\rm TE}=\cos\varphi_1 \cos\varphi_2 - \displaystyle\frac{ k_{1z}}{k_{2z}}\sin\varphi_1 \sin\varphi_2,
\end{equation}
\begin{equation}\label{eq18}
    \mathrm{M}_{12}^{\rm TE}=-i\displaystyle\frac{\omega}{c k_{1z}}\sin\varphi_1 \cos\varphi_2 - i\displaystyle\frac{\omega}{c k_{2z}}\sin\varphi_2 \cos\varphi_1,
\end{equation}
\begin{equation}\label{eq19}
    \mathrm{M}_{21}^{\rm TE}=-i\displaystyle\frac{c k_{1z}}{\omega}\sin\varphi_1 \cos\varphi_2 - i\displaystyle\frac{c k_{2z}}{\omega}\sin\varphi_2 \cos\varphi_1,
\end{equation}
\begin{equation}\label{eq20}
    \mathrm{M}_{22}^{\rm TE}=\cos\varphi_1 \cos\varphi_2 - \displaystyle\frac{k_{2z}}{k_{1z}}\sin\varphi_1 \sin\varphi_2.
\end{equation}
Afterwards, we obtain the dispersion relation for the TE surface wave in the form,
\begin{equation}\label{eq26}
    \sin\varphi=-\displaystyle\frac{2\pi i\sigma}{c} \mathrm{M}_{12}^{\rm TE},\
\end{equation}
where $\cos\varphi=(\mathrm{M}_{ 11}^{\rm TE}+\mathrm{M}_{22}^{\rm TE})/2$~\cite{Yeh}.

\section{Analysis of the dispersion curves}

For the numerical calculations of the dispersion curves, we assume that $T=10$~K and the graphene electron concentration $n=1.5\cdot 10^{15}$~m$^{-2}$. The corresponding chemical potential $\mu$ is about 45~meV (or 522~K). These parameters are in agreement with experimental results~\cite{Li}. We will neglect the dissipative losses and consider the graphene conductivities $\sigma^{\rm intra}$ and $\sigma^{\rm inter}$ as purely imaginary values.

\subsection{Dispersion curves for the TM-polarized surface modes}

\begin{figure}
\includegraphics [width=15.0 cm,height=11.0 cm]{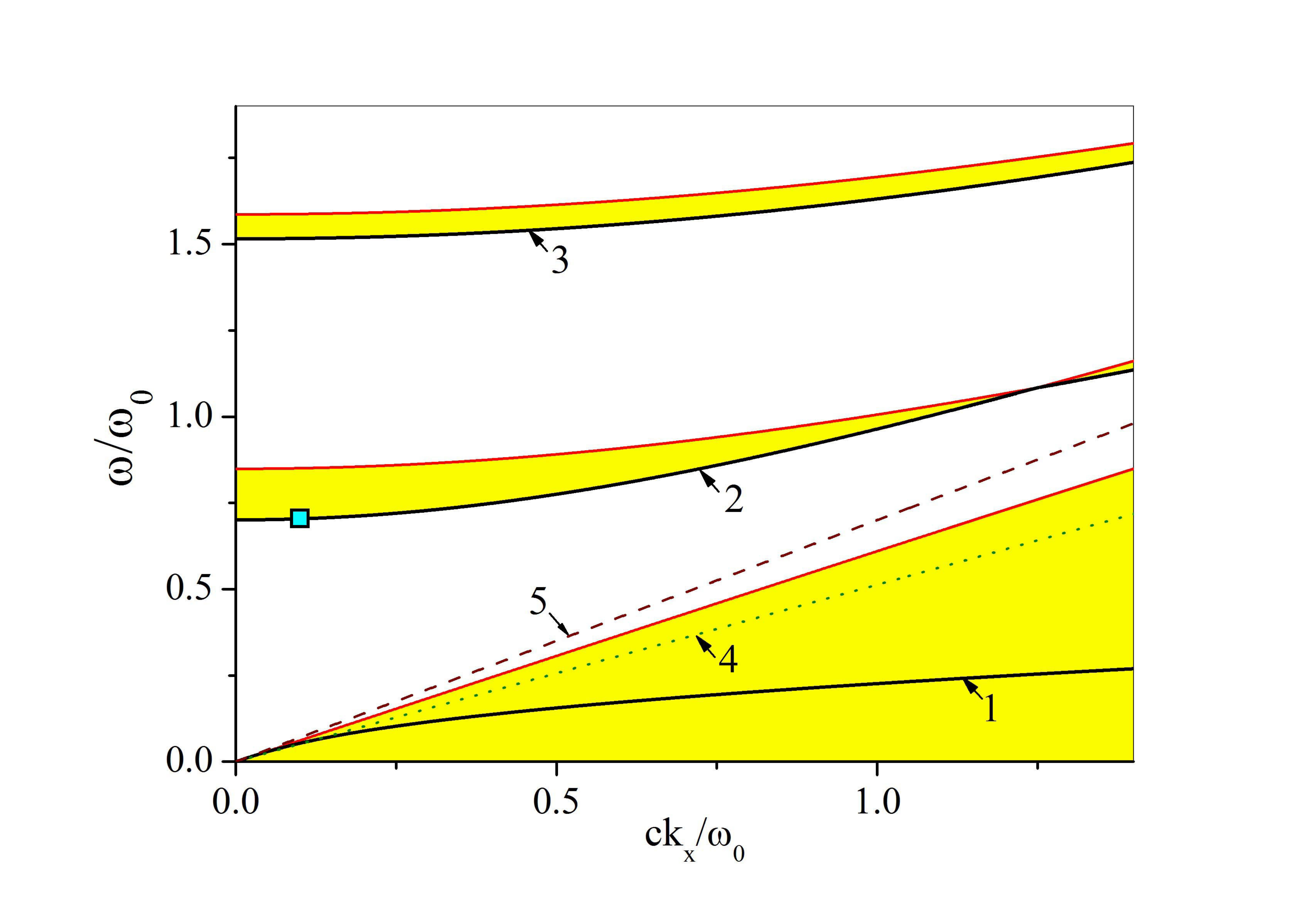}
\caption{\label{Fig2} (Color online) Dispersion curves (1--3) for the TM-polarized surface waves in the PC-graphene-PC structure in the THz frequency range. Curves 4 and 5 are the light lines for silica glass and Teflon, respectively. Here $\mu/\hbar\omega_0\approx 11$. The forbidden zones are marked in yellow. The square on curve 2 indicates the values of $\omega$ and $k_x$ which correspond to the peak of the transmissivity arising due to the excitation of the TM-polarized mode (see the next section).}
\end{figure}

Curves 1--3  in Fig.~\ref{Fig2} show the numerically-calculated dispersion curves for the TM-polarized surface waves. The wave frequency is normalized to $\omega_0=2\pi\cdot 10^{12}$~sec$^{-1}$ (terahertz range).
Hereafter, we assume the following parameters for the photonic crystal: $d_1=d_2=1.2\,c/\omega_0$, $\varepsilon_1=3.8$ (this value corresponds to silica glass), and $\varepsilon_2=2.04$ (Teflon) (see Refs.~\cite{Quartz,Teflon}). Curves 4 and 5 are the light lines for silica glass and Teflon, respectively.
 All the dispersion curves run within the forbidden zones (regions marked in yellow) of the infinite photonic crystal. Note that the edges of these forbidden zones, expressed in dimensionless coordinates $\omega/\omega_0$ and $c k_x/\omega_0$, are the same for different frequency ranges.  However, the positions of the dispersion curves depend substantially on the choice of $\omega_0$. One can see that each of the forbidden zones of the photonic crystal contains one branch of the spectrum for the TM-polarized surface wave.

 Note that the dispersion curve, which corresponds to the dispersion relation for the waves localized near the graphene layer placed between the semi-infinite Teflon and semi-infinite silica glass, practically coincides with curve 1 in Fig.~\ref{Fig2}.
This coincidence means that replacing the semi-infinite Teflon and semi-infinite silica glass by the photonic crystals does not change significantly the dispersion curve for the TM-polarized surface wave in the THz frequency range. However, this  remark only concerns the first forbidden zone.

This fact has a simple physical interpretation. Namely, the plasmons (TM modes) near the graphene layer are extremely localized and, at large frequencies, the evanescent tail can be smaller than the thickness of the dielectric layers. In other words, graphene can truly ``see'' only the nearest uniform dielectrics. On the other hand, in the low-frequency limit (when the evanescent tail becomes larger), the graphene layer will be able to ``see'' many layers of the photonic crystal, changing the plasmon dispersion relation. To demonstrate this, we present some numerical estimates for the dimensionless localization depth $\xi=k_x/\mathrm{Im}(q)$ of the electromagnetic fields of the TM-polarized waves. Recall that $k_x$ is the wave vector of the localized mode and $\mathrm{Im}(q)$ is the Bloch decay constant. Physically, the value of $\xi$ is the ratio of the distance from the graphene layer, where the field amplitudes are reduced by a factor of e~$\approx 2.718$, to the wavelength. For the mode in the first forbidden zone, we have: $\xi\approx 1.04$ in the THz frequency range and $\xi\approx 4.35$ in the centimeter-wavelength frequency range (for $\omega_0=2\pi\cdot 10^{10}$~sec$^{-1}$). Note that curves 2 and 3 on Fig.~\ref{Fig2} seem to be lying precisely at the boundaries of the forbidden regions (marked in yellow). This means that these modes are  weakly localized near the graphene layer. Indeed, we have $\xi\approx 534$, and  $\xi\approx 176$ for the modes in the second and third forbidden zones, respectively.  The main new remarkable feature of the spectrum for the surface waves in the PC-graphene-PC system, in comparison with the Teflon-graphene-silica glass structure, consists in the appearance of additional branches of the spectrum in the second, third, and so on, forbidden zones.

\subsection{Dispersion curves for the TE-polarized surface modes}

\begin{figure}
\includegraphics [width=15.0 cm,height=11.0 cm]{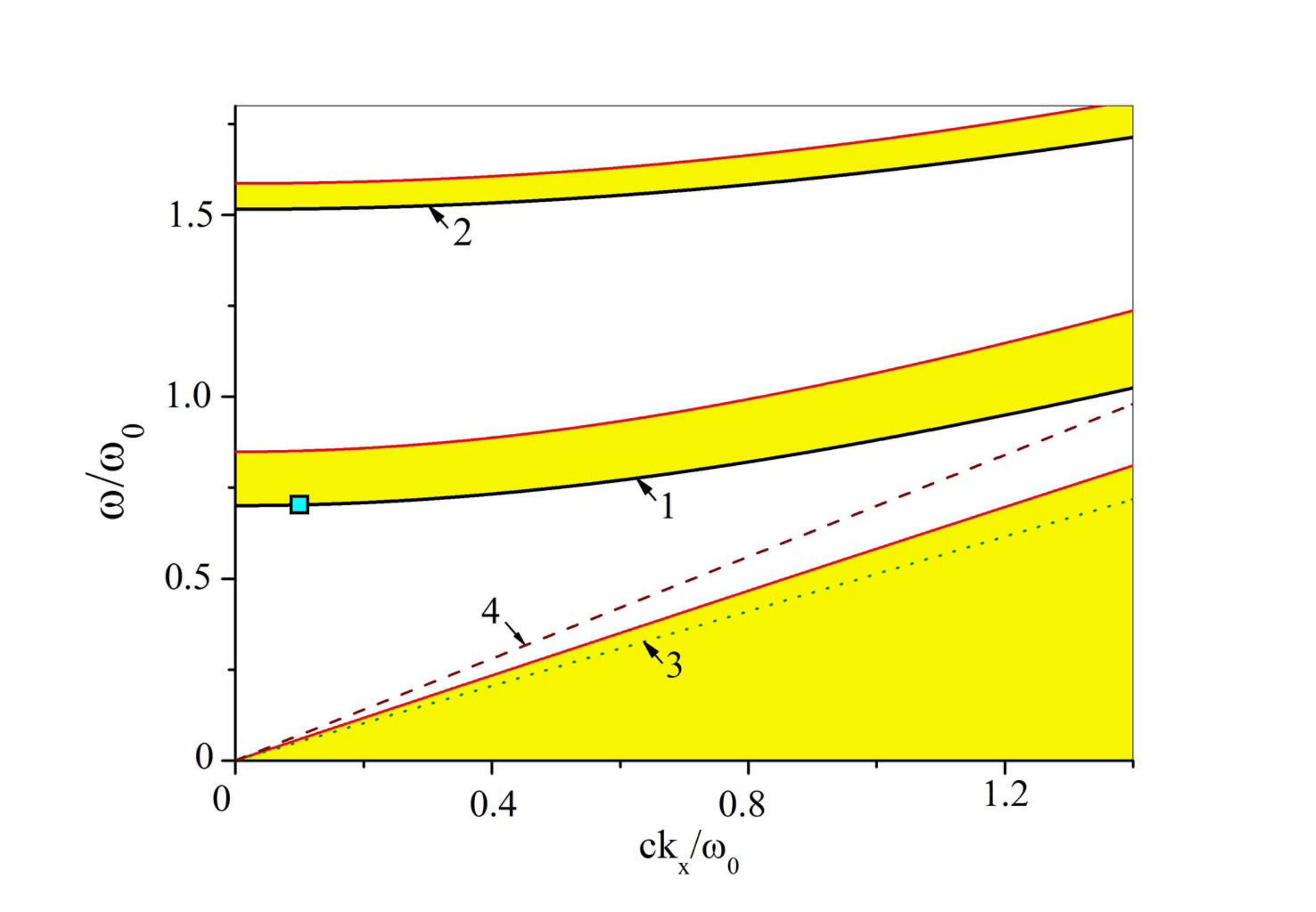}
\caption{\label{Fig3} (Color online) Dispersion curves for the TE-polarized surface waves in the structure PC-graphene-PC in the THz frequency range (curves 1 and 2).  Curves 3 and 4 are the light lines for silica glass and Teflon, respectively. Here $\mu/\hbar\omega_0\approx 11$. The forbidden zones are marked in yellow. The square on curve 1 indicates the values of $\omega$ and $k_x$ which correspond to the peak of the transmissivity arising due to the excitation of the TE-polarized mode (see the next section).}
\end{figure}

Curves $1$, $2$ in Fig.~\ref{Fig3} show the dispersion curves for the TE-polarized surface waves in the THz frequency range, for the same parameters of the photonic crystal as in Fig.~\ref{Fig2}. Curves 3 and 4 in Fig.~\ref{Fig3} are the light lines for silica glass and Teflon, respectively. The principal difference of the spectrum for the TE-polarized waves, in comparison with TM waves, is the absence of dispersion curves in the first forbidden zone for the terahertz range. Indeed, the dispersion curve appears in the first forbidden zone only when $\mathrm{Im}(\sigma)<0$. In contrast to the usual two-dimensional electron gas with Drude conductivity, this inequality can be satisfied in monolayer (or bilayer~\cite{bilayer}) graphene. For the chosen values of $n$ and $T$, this condition is only satisfied starting with the infrared (IR) frequencies. As seen from Fig.~\ref{Fig3}, the dispersion curves for the TE modes, as well as for the TM modes (see the dispersion curves $2$ and $3$ in Fig.~\ref{Fig2}), are very close to the boundaries of the forbidden zones. This means that these modes are weakly localized near the graphene layer. The numerical estimates of the dimensionless localization depth $\xi$ for the TE waves give similar values as for the corresponding TM-polarized modes.

\begin{figure}
\includegraphics [width=15.0 cm, height=11.0 cm]{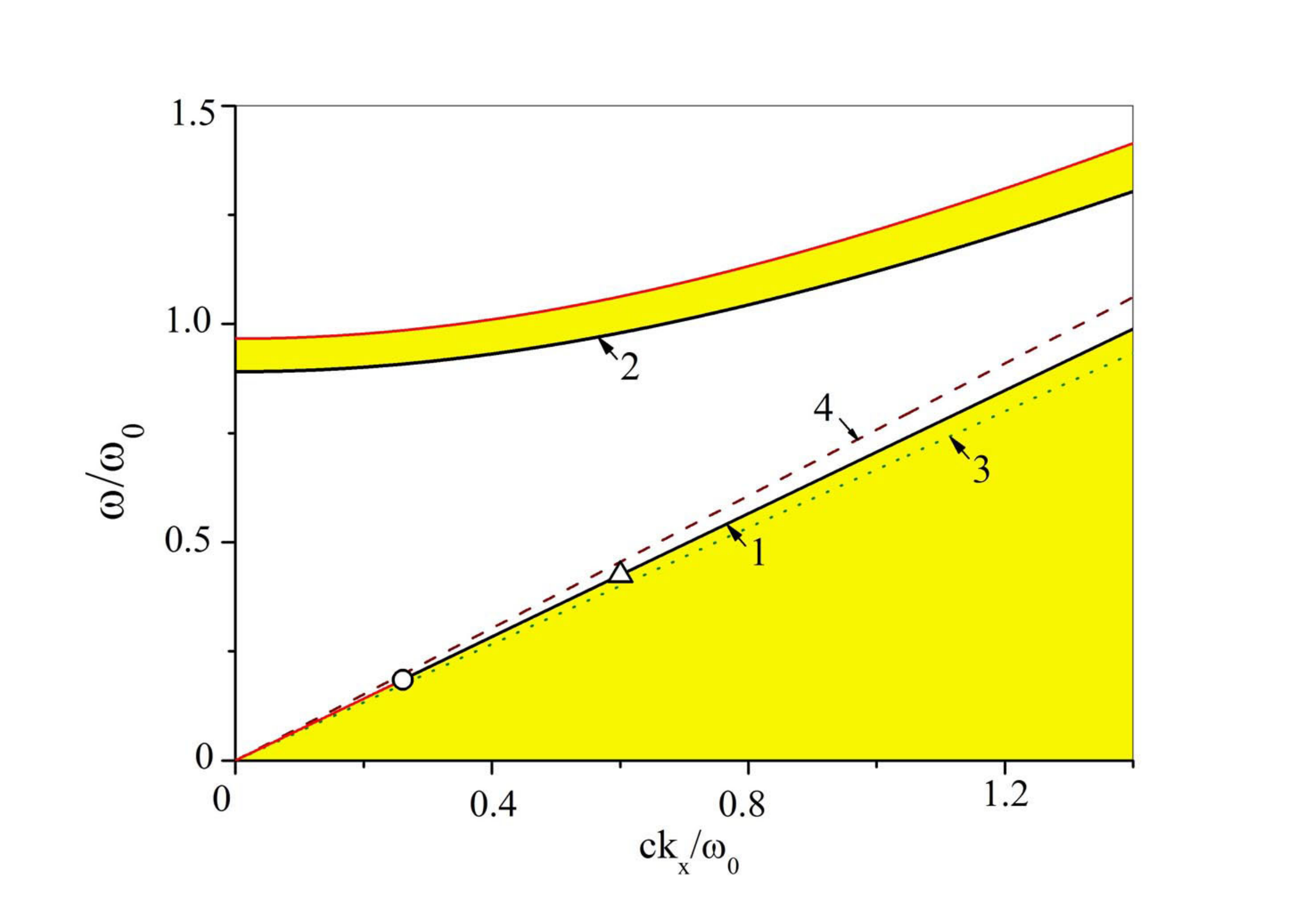}
\caption{\label{Fig4} (Color online) Dispersion curves for the TE-polarized surface waves in the structure PC-graphene-PC in the IR frequency range (curves $1$ and $2$). Curves 3 and 4 are the light lines for silica glass and Teflon, respectively. Here $\mu/\hbar\omega_0\approx 0.11$. The forbidden zones are marked in yellow. The open circle indicates the starting point of dispersion curve 1. The part of curve $1$ between the circle and triangle corresponds to the highest degree of localization of the mode's field.}
\end{figure}

\begin{figure}
\includegraphics [width=15.0 cm, height=11.0 cm]{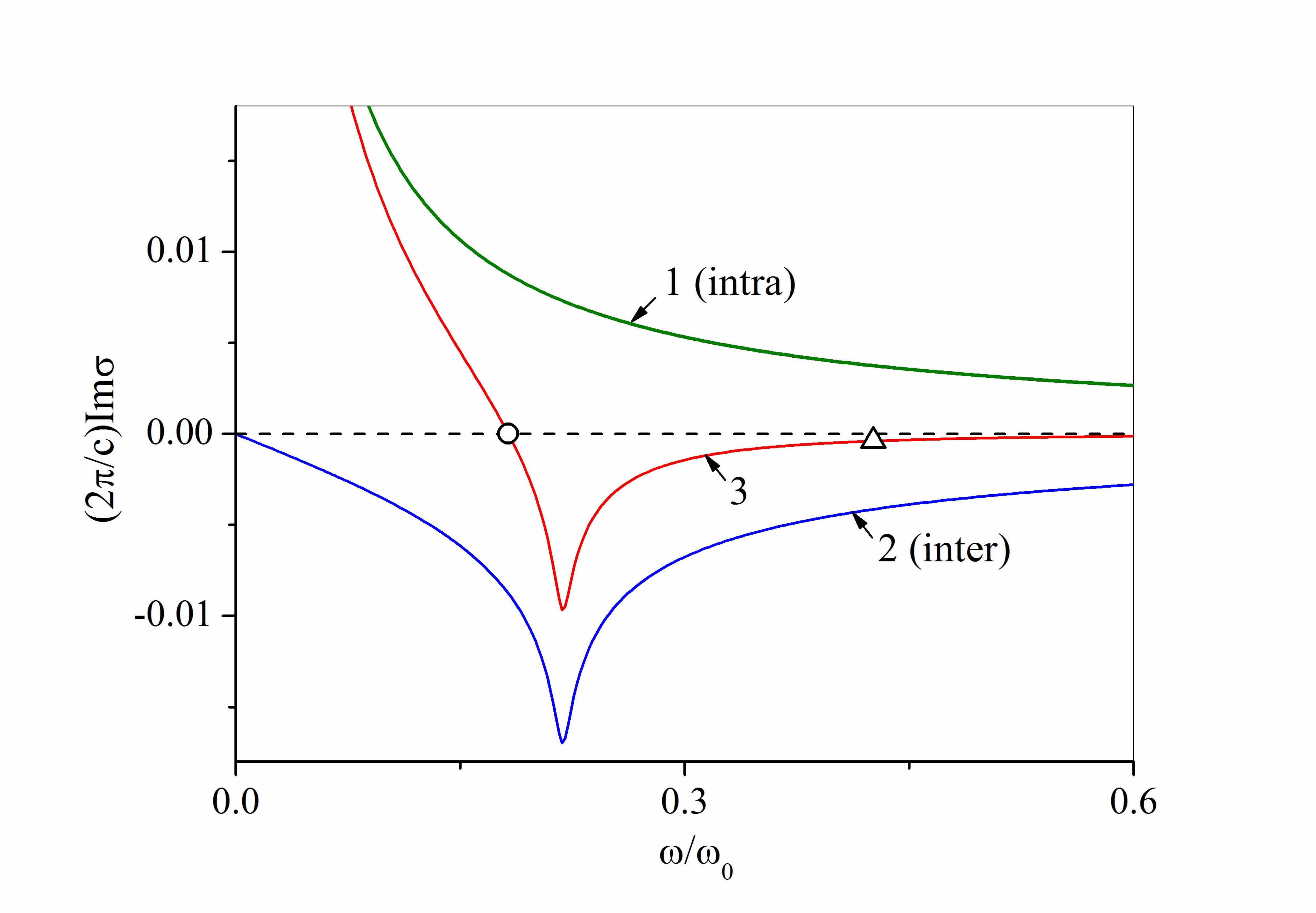}
\caption{\label{Fig5} (Color online) Frequency dependence of the dimensionless conductivities $(2\pi/c)\mathrm{Im}(\sigma^{\rm intra})$ (curve 1), $(2\pi/c)\mathrm{Im}(\sigma^{\rm inter})$ (curve 2), and $(2\pi/c)\mathrm{Im}(\sigma)$ (curve 3) calculated by means of Eqs.~\eqref{eq10} and \eqref{eq11} for the given values of $\mu=522$~K, $T=10$~K, and $\omega_0=2\pi\cdot 10^{14}$~sec$^{-1}$, at $\nu=0$.}
\end{figure}

Now we consider the dispersion curves of the TE-polarized modes in the IR frequency range, where unusual transport properties of the graphene layer manifest themselves.
These curves (1 and 2) are shown in Fig.~\ref{Fig4} for the same parameters of the photonic crystal as in Fig.~\ref{Fig3} (however, with the dielectric permittivities $\varepsilon_1=2.25$ and $\varepsilon_2=1.74$ for silica glass and Teflon in the IR frequency range, see Refs.~\cite{Quartz,Teflon}). The wave frequency is normalized to $\omega_0=2\pi\cdot 10^{14}$~sec$^{-1}$.
As seen from Fig.~\ref{Fig4}, dispersion curve 2 for the TE-polarized mode in the second forbidden zone is very close to the boundary of the forbidden zone, similarly to the case of the THz frequency range. This means that this mode is also weakly localized near the graphene layer. The dispersion curve for the TE mode in the first forbidden zone exists only in a finite-frequency interval. The edges of the corresponding part of the dispersion curve are depicted by the open circle and open triangle on curve 1 in Fig.~\ref{Fig4}. At higher frequencies, the Bloch phase $\mathrm{Im}(\varphi)$ in the dispersion relation given by Eq.~\eqref{eq26} is very small,  $\mathrm{Im}(\varphi) < 10^{-5}$. The localization depth is much larger than the wavelength of the TE mode in this case. To clarify this, it is worthwhile to analyze the frequency-dependence of the conductivity, $\mathrm{Im}(\sigma)=\mathrm{Im}(\sigma^{\rm intra})+\mathrm{Im}(\sigma^{\rm inter})$, for the given values of $\mu$, and $T$, at $\nu=0$. This dependence is shown in Fig.~\ref{Fig5} for the IR frequency range. The open circle and open triangle in Fig.~\ref{Fig5} are in agreement with the same symbols shown on curve $1$ in Fig.~\ref{Fig4}.
The maximum absolute value of $\mathrm{Im}(\sigma)$ corresponds to the maximum of the Bloch phase $\mathrm{Im}(\varphi)\sim 5\cdot 10^{-3}$ and to the strongest localization of the electromagnetic field of the TE wave near the graphene layer. The conductivity $\mathrm{Im}(\sigma)$ tends to zero when increasing the frequency. As a result, the localization depth becomes 100 times larger for the point marked by the open triangle on curve $1$ in Fig.~\ref{Fig4}.

\subsection{Dispersion curves for TE-polarized surface modes versus the PC parameters}

Note that there are three necessary conditions for the existence of TE-polarized surface waves propagating along a conducting layer placed between \emph{semi-infinite homogeneous} dielectrics. First, according to Ref.~\onlinecite{Mikhailov}, the linear electron-hole dispersion law, specific for the graphene layer, should be valid. Second, the condition $\mathrm{Im}(\sigma)<0$ should be satisfied. Finally, the permittivities of the two dielectrics surrounding the graphene layer should be \emph{very similar}. Here we now consider the system PC-graphene-PC where the graphene layer is placed between two \emph{very different} adjacent dielectrics. So, at first glance, the localized TE modes cannot exist in such a system because the third condition mentioned above is not satisfied. However, the dimensionless localization depth $\xi$ for all the TE surface modes considered are extremely large, and the graphene layer can actually ``see'' a symmetric dielectric environment which enables the existence of the TE modes.

Here we present results of our numerical analysis of the dispersion curve for the TE localized mode in the lowest forbidden zone of the photonic crystal versus the thicknesses $d_1$ and $d_2$ of the dielectric layers. In the case $d_1>d_2$, i.e., when the thicknesses of the silica glass layers are greater than the thicknesses of the Teflon layers, the boundary of the first forbidden zone of the photonic  crystal in Fig.~\ref{Fig4}, as well as the dispersion curve for the TE-polarized mode (curve $1$ in ~\ref{Fig4}) shift towards the light line in silica glass (curve 3 in Fig.~\ref{Fig4}). When the inverse inequality is realized, i.e., when $d_1<d_2$, the boundary of the first forbidden zone and the dispersion curve for the TE-polarized mode shift towards the light line in Teflon (curve 4 in Fig.~\ref{Fig4}). In addition, the dispersion curve for the TE-polarized mode shifts towards the light line in silica glass when increasing the thicknesses of the PC layers (for $d_1=d_2\gg 1$ ) and tends to this light line in the limit $d_1=d_2\rightarrow\infty$. This means that, in the limit $d_1=d_2\rightarrow\infty$, the electromagnetic field of the localized TE wave becomes delocalized in silica glass, and the localized wave itself does not exist.

In this section, we used the Floquet theorem \cite{Morse-Feshbach} to derive the dispersion relations for the TM- and TE-polarized eigenmodes localized near the graphene layer inserted into a photonic crystal with an infinite number of layers. As known, this theorem is valid for the non-dissipative case only.  However, the behavior of the dispersion curves does not change significantly for systems with small enough dissipative parameters. We demonstrate this in the next section where we study the problem of the excitation of localized waves in the PC-graphene-PC structure with a finite number of layers. We will not use the Floquet theorem, but instead apply the propagation-matrix method, which allows us to take into account realistic dissipation parameters.

\section{Excitation of the TE- and TM-polarized surface waves}

\subsection{Transmissivity of the PC-graphene-PC structure with a finite number of layers}

In this section, we study the transmittance of the PC-graphene-PC structure for TE-polarized waves. The calculation of the transmission coefficient for the TM-polarized waves is performed in the same way. We now consider the excitation of the TE waves propagating along the graphene layer placed in the middle of the photonic crystal containing a \emph{finite number} $2N$ of elementary cells. This structure is irradiated, by an electromagnetic wave, from its right side. The external wave is incident from a dielectric with permittivity $\varepsilon_d$ on the PC-graphene-PC structure under the angle $\arcsin (k_x c /\omega \sqrt{\varepsilon_d}) $, passes through the PC-graphene-PC structure, and, finally, transmits to another external dielectric with the same permittivity $\varepsilon_d$. Evidently, the resonance excitation of the surface wave will occur if its polarization, frequency, and the component $k_x$ of the wave vector coincide with the parameters of the incident wave.

We define the electric and magnetic fields in the dielectric on the right-hand side of the structure (at $z>2Nd$) as
\begin{eqnarray}
 & E_y=A_0\exp\bigl\{-i k_{zd}(z-2Nd)\bigr\}\nonumber\\
 & + B_0\exp\bigl\{i k_{zd}(z-2Nd)\bigr\},\
\end{eqnarray}
\begin{eqnarray}
 & H_x=\displaystyle\frac{c k_{zd}}{\omega}\Bigl[A_0\exp\bigl\{-i k_{zd}(z-2Nd)\bigr\} \nonumber\\
 & - B_0\exp\bigl\{i k_{zd}(z-2Nd)\bigr\}\Bigr],\
\end{eqnarray}
where $A_0$ and $B_0$ are the amplitudes of the incident and reflected waves, respectively. We define the electromagnetic field of the wave on the left-hand side of the structure (at $z<0$) as
\begin{equation}\label{EqS3}
    E_y=C_0\exp(-i k_{zd} z),\ H_x=\displaystyle\frac{c k_{zd}}{\omega}C_0\exp(-i k_{zd} z),\
\end{equation}
where $C_0$ is the amplitude of the transmitted wave.

\begin{figure}
\includegraphics [width=15.0 cm, height=11.0 cm]{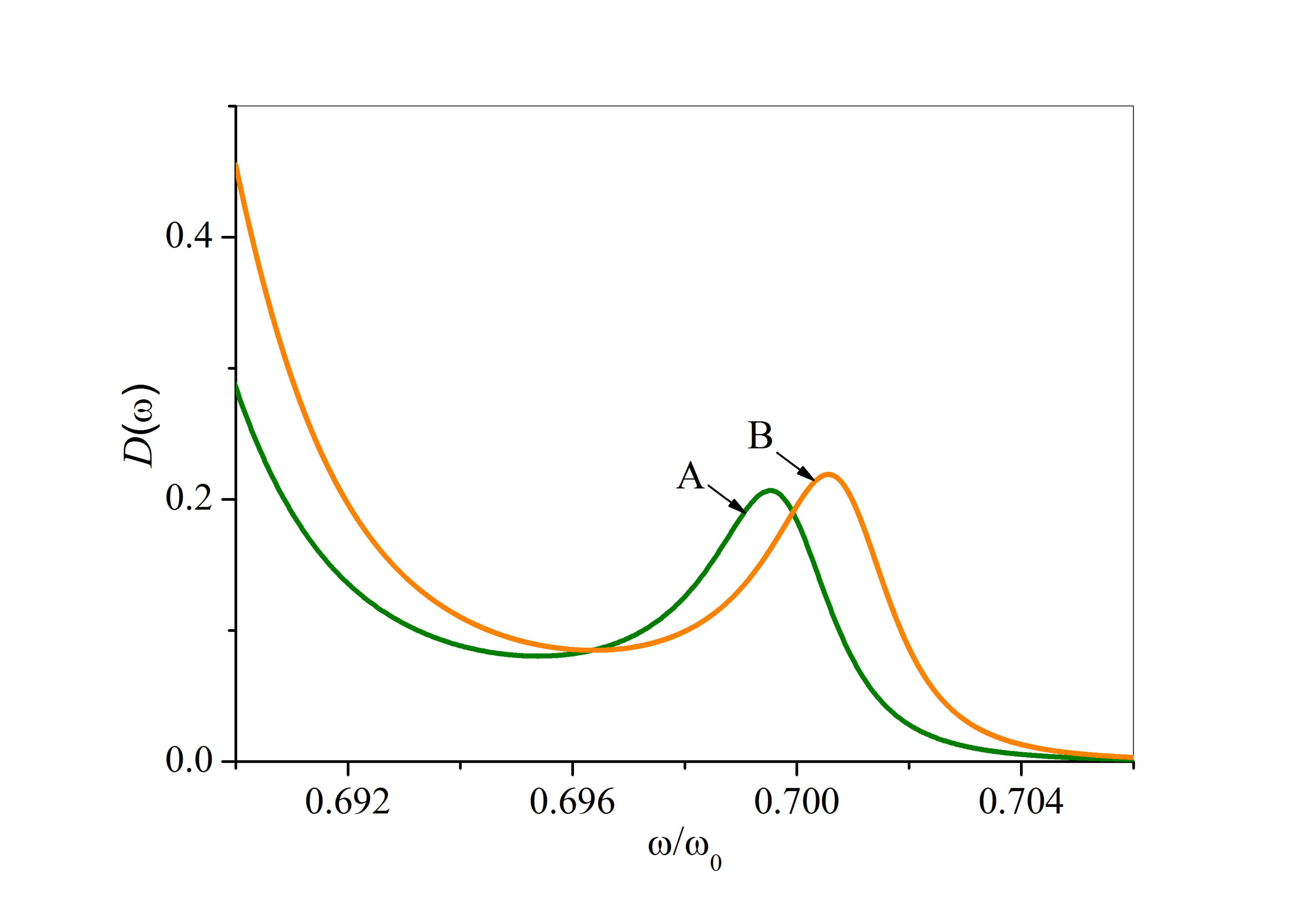}
\caption{\label{Fig6} (Color online) Frequency dependence of the transmissivity $D(\omega)$ for the PC-graphene-PC structure for the TE (curve A) and TM (curve B) waves. The resonance frequencies and the values of other parameters correspond to the squares on the dispersion curves in Figs. 2 and 3. Note that we have $k_{zd}/k_x\approx 6.92$ and $k_{zd}/k_x\approx 6.93$ for the maxima on curves A and B, respectively.}
\end{figure}

To find the transmission coefficient $t=C_0/A_0$, we use the condition of the magnetic field discontinuity on a graphene layer (similar to Eq.~\eqref{eq9b} written for the TM wave) and Eq.~\eqref{eq15}. Following the procedure described in Ref.~\cite{Yeh}, we derive,
 \begin{equation}\label{EqS4}
      \begin{pmatrix} E_y(z=2Nd)  \\ H_x(z=2Nd) \end{pmatrix}=\mathrm{\bf P}^{\rm TE} \begin{pmatrix} E_y(z=0)  \\ H_x(z=0) \end{pmatrix}\
 \end{equation}
with $\mathrm{\bf P}^{\rm TE}=\bigl(\mathrm{\bf M}^{\rm TE}\bigr)^N \mathrm{\bf M}_{\rm Gr}^{\rm TE} \bigl(\mathrm{\bf M}^{\rm TE}\bigr)^N$.
Substituting Eqs.~(18)--(\ref{EqS3}) into Eq.~\eqref{EqS4}, we derive the following relation for the wave amplitudes:
\begin{equation}\label{EqS5}
    \begin{pmatrix} A_0+B_0  \\ \displaystyle\frac{c k_{zd}}{\omega} (A_0-B_0) \end{pmatrix}=\mathrm{\bf P}^{\rm TE} \begin{pmatrix} C_0  \\ \displaystyle\frac{c k_{zd}}{\omega}C_0 \end{pmatrix}.
\end{equation}
Solving Eq.~\eqref{EqS5}, we have the following expression for the Fresnel transmissivity coefficient:
\begin{equation}\label{eq33}
    t=\displaystyle\frac{2}{P^{\rm TE}_{11}+P^{\rm TE}_{22}+\displaystyle\frac{\omega}{c k_{z d}}P^{\rm TE}_{21}+\displaystyle\frac{c k_{z d}}{\omega}P^{\rm TE}_{12}}\
\end{equation}
with $k_{z d}=\sqrt{\varepsilon_d k_0^2-k_x^2}$. The corresponding expression for the TM-polarized wave is
\begin{equation}\label{eq31}
    t=\displaystyle\frac{2}{P^{\rm TM}_{11}+P^{\rm TM}_{22}-\displaystyle\frac{\omega\varepsilon_d}{c k_{zd}}P^{\rm TM}_{21}-\displaystyle\frac{c k_{z d}}{\omega\varepsilon_d}P^{\rm TM}_{12}}.
\end{equation}
Here $P^{\rm TM}_{jk}$ and $P^{\rm TE}_{jk}$ are the elements of the matrices  $\mathrm{\bf P}^{\rm TE}$ and $\mathrm{\bf P}^{\rm TM}=\bigl(\mathrm{\bf M}^{\rm TM}\bigr)^N \mathrm{\bf M}_{\rm Gr}^{\rm TM} \bigl(\mathrm{\bf M}^{\rm TM}\bigr)^N$. The matrices $\mathrm{\bf M}_{\rm Gr}^{\rm TE}$ and $\mathrm{\bf M}_{\rm Gr}^{\rm TM}$ are,
\begin{equation}\label{MGrTM}
      \mathrm{\bf M}_{\rm Gr}^{\rm TE}=\begin{pmatrix} 1 & 0 \\ 4\pi \sigma/c & 1 \end{pmatrix}, \quad
      \mathrm{\bf M}_{\rm Gr}^{\rm TM}=\begin{pmatrix} 1 & -4\pi \sigma/c \\ 0 & 1 \end{pmatrix}.
\end{equation}
We now study the frequency dependence $D(\omega)=|t(\omega)|^2$ of the transmissivity of the PC-graphene-PC structure, which should have a sharp maximum when the resonance excitation of the surface wave takes place. Figure~\ref{Fig6} shows the frequency dependence  $D(\omega)$ for the TM- and TE-polarized waves at $\omega_0=2\pi\cdot 10^{12}$~sec$^{-1}$, $k_x c/\omega_0=0.1$, $N=15$, $\varepsilon_d=1$ (vacuum), and $\nu=2\cdot 10^{12}$~sec$^{-1}$. The chosen value of the relaxation frequency is in agreement with the theoretical and experimental results~\cite{Hwang,17,Li}.
The resonance peaks of the transmissivity are observed within the forbidden zone at frequencies that correspond to the open squares on the dispersion curves in Figs.~\ref{Fig2} and \ref{Fig3}. Note that these peaks are clearly seen in spite of the weak localization of the excited modes. The positions of the peaks, that can be measured for any wave polarization, and the quite-simple Eqs.~\eqref{eq14} and \eqref{eq26} allow one to calculate the graphene conductivity in the centimeter, millimeter, and submillimeter wavelength frequency ranges. Note also that, in the absence of the graphene layer in PC, the peaks of the transmissivity are not observed.

\subsection{Spatial distributions of the squared electromagnetic field amplitudes for the excited localized waves}

\begin{figure}
\includegraphics [width=15.0 cm,height=11.0 cm]{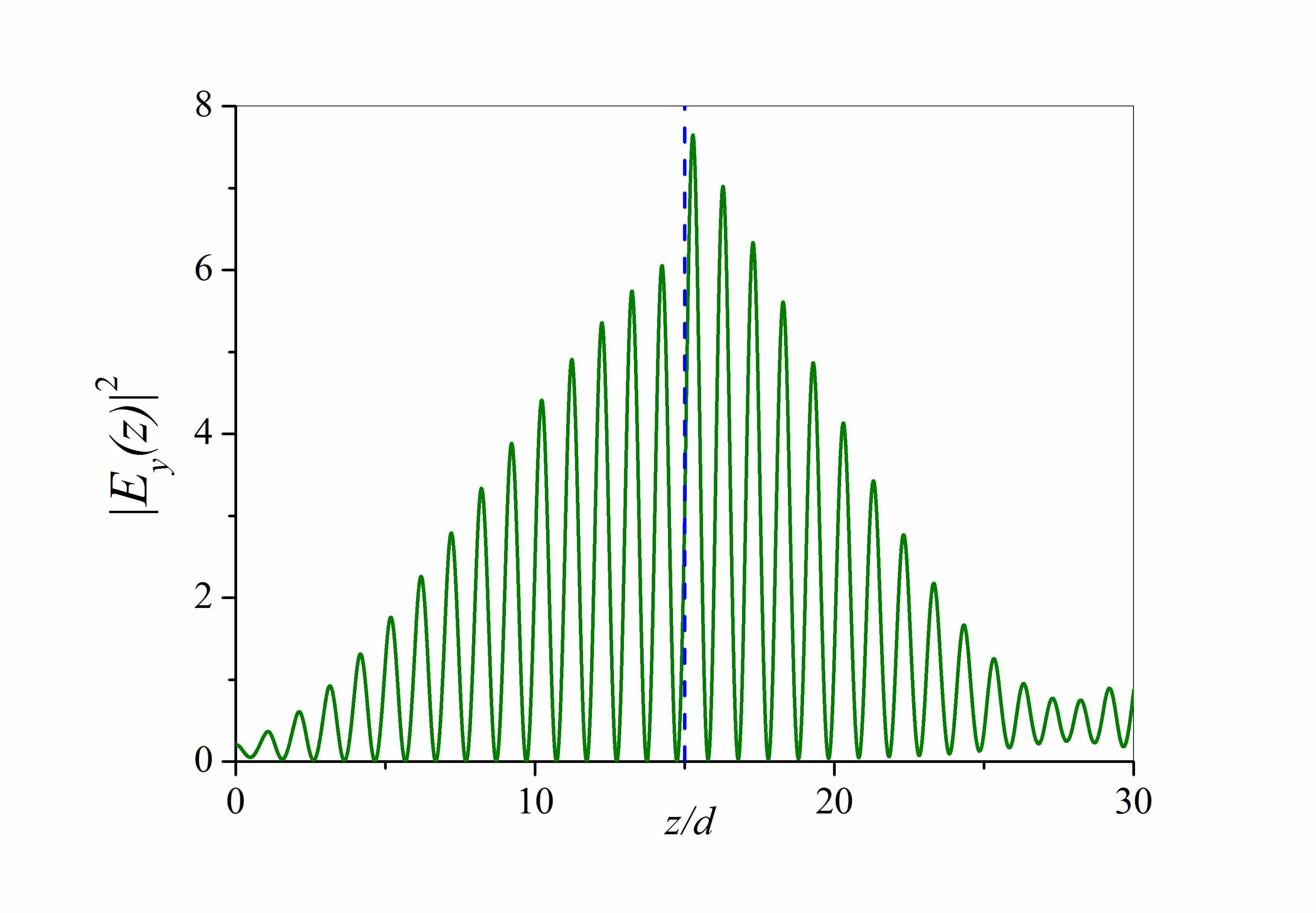}
\caption{\label{Fig7} (Color online) The distribution of the squared amplitude $|E_y(z)|^2$ of the dimensionless electric field for the TE-polarized wave $E_y(z)$ in the PC-graphene-PC structure.}
\end{figure}

\begin{figure}
\includegraphics [width=15.0 cm,height=11.0 cm]{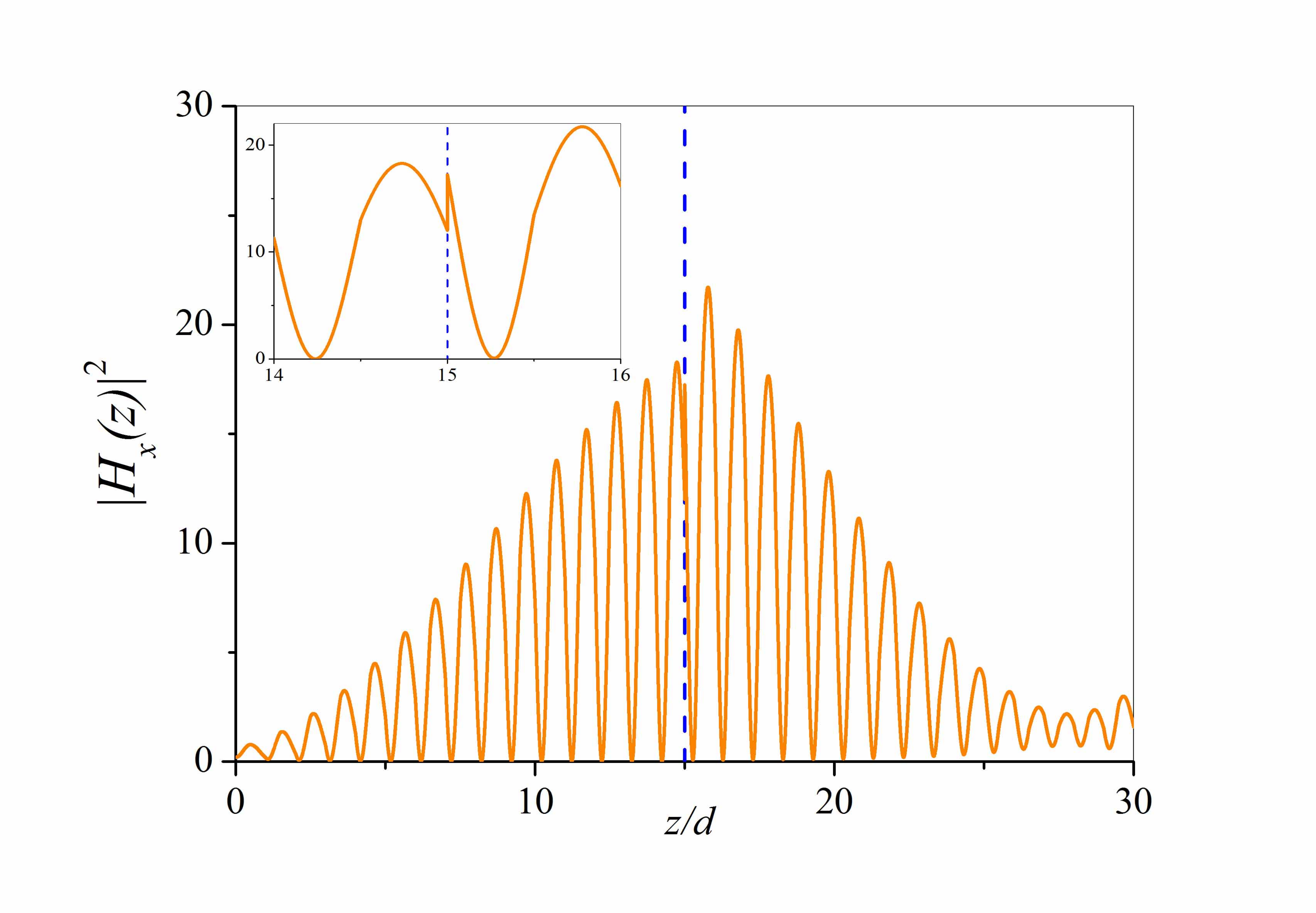}
\caption{\label{Fig8} (Color online) The distribution of the squared amplitude $|H_x(z)|^2$ of the dimensionless magnetic field for the TE-polarized wave $H_x(z)$ in the PC-graphene-PC structure.}
\end{figure}

The presence of a graphene layer within a photonic crystal results in a concentration of the electromagnetic field near it. In this section, we discuss the spatial distribution of the squared amplitudes $|E_y(z)|^2$ and $|H_x(z)|^2$ of the electric and magnetic fields in the problem of the excitation of the localized wave by the incident electromagnetic wave (see the previous section). The distributions of $|E_x(z)|^2$ and $|H_y(z)|^2$ for the TM-polarized wave are qualitatively the same as the distributions $|E_y(z)|^2$ and $|H_x(z)|^2$ for the TE wave.
For numerical calculations, we chose the frequency which corresponds to the maximum of the peak on curve A in Fig.~\ref{Fig6}. The results of the calculations are shown in Figs.~\ref{Fig7} and ~\ref{Fig8}. The values $|E_y(z)|^2$ and $|H_x(z)|^2$ are normalized to the corresponding values of the incident wave. The dashed vertical lines in the figures show the position of the graphene layer. We assume that the light is incident from the right-hand side of the structure, i.e., the wave propagates from the unit cell with number 30 to the unit cell with number 1.
Figures~\ref{Fig7} and \ref{Fig8} show that the maxima in the distributions of the electric and magnetic fields are near the graphene plane. In addition, the value of $|H_x(z)|^2$ suffers a discontinuity on the graphene plane (see the inset in Fig.~\ref{Fig8}) which is caused by the ac current excited in the graphene layer by the incident wave.

\section{Conclusion}

To conclude, we predict the coexistence of localized TE- and TM-polarized surface electromagnetic waves in a PC-graphene-PC structure. These waves can propagate \emph{in the same frequency range} due to the zone structure of the photonic crystal, in contrast to the waves localized near the graphene layer placed between two identical dielectrics. In the latter case, the TM and TE surface waves can propagate in different frequency ranges. We also consider the  excitation of localized TM and TE modes by the external wave that irradiates the PC-graphene-PC structure. We show that, independently of the polarization of the exciting wave, the resonance peak of the transmissivity of the structure appears when changing the frequency or the incident angle. The analysis of the resonance peaks of the wave transmissivity can provide important information on the graphene conductivity in the centimeter, millimeter, and submillimeter wavelength frequency ranges.

\section{Acknowledgements}

We acknowledge partial support from the RIKEN iTHES project, JSPS-RFBR Contract No.~12-02-92100, Grant-in-Aid for Scientific Research (S), and the Program FPNNN of the NAS of Ukraine (Project No.~0110U005642).


\begin{references}

\bibitem{HE} K.S. Novoselov, A.K. Geim, S.V. Morozov, D. Jiang, M.I. Katsnelson, I.V. Grigorieva, S.V. Dubonos, and A.A. Firsov, Nature \textbf{438}, 197 (2005); Y. Zhang,  Y.-W. Tan, H.L. Stormer, and P. Kim, Nature \textbf{438}, 201 (2005).

\bibitem{KP} M.I. Katsnelson, K.S. Novoselov, and A.K. Geim, Nature Physics \textbf{2}, 620 (2006).

\bibitem{AB} P. Recher, B. Trauzettel, A. Rycerz, Ya.M. Blanter, C.W.J. Beenakker, and A.F. Morpurgo,  Phys. Rev. B. \textbf{76}, 235404 (2007).

\bibitem{WL} S.V. Morozov, K.S. Novoselov, M.I. Katsnelson, F. Schedin, L.A. Ponomarenko, D. Jiang, and A.K. Geim, Phys. Rev. Lett. \textbf{97}, 016801 (2006).

\bibitem{CF} C. Berger, Z. Song, X. Li, X. Wu, N. Brown, C. Naud, D. Mayou, T. Li, J. Hass, A.N. Marchenkov, E.H. Conrad, P.N. First, and W.A. de Heer, Science \textbf{312}, 1191 (2006).

\bibitem{QN} R. Danneau, F. Wu, M.F. Craciun, S. Russo, M.Y. Tomi, J. Salmilehto, A.F. Morpurgo, and P.J. Hakonen, Phys. Rev. Lett. \textbf{100}, 196802 (2008).

\bibitem{CB} F. Sols, F. Guinea, and A.H. Castro Neto,  Phys. Rev. Lett. \textbf{99}, 166803 (2007).

\bibitem{AL} A.H. Castro Neto, F. Guinea, and N.M.R. Peres, Phys. Rev. B. \textbf{73}, 205408 (2006).

\bibitem{AR} C.W.J. Beenakker, Phys. Rev. Lett. \textbf{97}, 067007 (2006); M. Titov and C.W.J. Beenakker, Phys. Rev. B \textbf{74}, 041401(R) (2006).

\bibitem{WC} C.-H. Zhang and Y.N. Joglekar, Phys. Rev. B. \textbf{75}, 245414 (2007).

\bibitem{VD} V.A. Yampol'skii, S. Savel'ev, and F. Nori, New J. Phys. \textbf{10}, 053024 (2008); V.A. Yampol'skii, S.S. Apostolov, Z.A. Maizelis, A. Levchenko, and F. Nori, Europhys. Lett. \textbf{96} 67009 (2011).

\bibitem{EL} V.V. Cheianov, V. Fal'ko, and B.L. Altshuler, Science \textbf{315}, 1252 (2007).

\bibitem{4} A.H. Castro Neto, F. Guinea, N.M.R. Peres, K.S. Novoselov,  and A.K. Geim, Rev. Mod. Phys. \textbf{81}, 109 (2009).

\bibitem{9a} A.V. Rozhkov, G. Giavaras, Yu.P. Bliokh, V. Freilikher,  and F. Nori, Phys. Rep. \textbf{503}, 77 (2011).

\bibitem{1} V.P. Gusynin , S.G. Sharapov, and J.P. Carbotte, Int. J. Mod. Phys. B, \textbf{21}, 4611(2007).

\bibitem{2} A.K. Geim  and K.S. Novoselov, Nat. Mater. \textbf{6}, 183 (2007).

\bibitem{3} C.W.J. Beenakker, Rev. Mod. Phys. \textbf{80}, 1337 (2008).

\bibitem{5} J.R.Williams, L. DiCarlo,  and C.M. Marcus, Science \textbf{317}, 638 (2007).

\bibitem{6} M. Fogler,  D.S. Novikov, L.I. Glazman, and B.I. Shklovskii, Phys. Rev. B \textbf{77}, 075420 (2008).

\bibitem{7} Yu.P. Bliokh, V. Freilikher, S. Savel'ev, and F. Nori, Phys. Rev. B \textbf{79}, 075123 (2009).

\bibitem{8} A.V.Rozhkov, S. Savel'ev, and F. Nori, Phys. Rev. B \textbf{79}, 125420 (2009).

\bibitem{9} N. Levy, S.A. Burke, K.L. Meaker, M. Panlasigui, A. Zettl, F. Guinea, A.H. Castro Neto, and M.F. Crommie,  Science \textbf{329}, 544 (2010).

\bibitem{Mikhailov} S.A. Mikhailov and K. Ziegler,  \prl \textbf{99}, 016803 (2007).

 \bibitem{10} E.H. Hwang and S. Das Sarma, \prb \textbf{80}, 205405 (2009).

\bibitem{11} E.G. Mishchenko, A.V. Shytov, and P.G. Silvestrov, \prl \textbf{104}, 156806 (2010).

 \bibitem{12} B. Wang, X. Zhang, F.J. Garc\'{\i}a-Vidal,  X. Yuan, and J. Teng,  \prl \textbf{109}, 073901 (2012).

 \bibitem{13} G.W. Hanson, E. Forati, W. Linz, and A.B. Yakovlev,  \prb \textbf{86}, 235440 (2012).

\bibitem{14}P.A. Huidobro, A.Y. Nikitin, C. Gonz\'{a}lez-Ballestero, L. Mart\'{\i}n-Moreno, and F.J. Garc\'{\i}a-Vidal,  \prb \textbf{85}, 155438 (2012).

\bibitem{15} D.A. Smirnova, A.V. Gorbach, I.V. Iorsh, I.V. Shadrivov, and Yu.S. Kivshar, \prb \textbf{88}, 045443 (2013).

\bibitem{16} A.V. Gorbach, Phys. Rev. A \textbf{87}, 013830 (2013).

\bibitem{17} I.V. Iorsh, I.V. Shadrivov, P.A. Belov, and Yu.S. Kivshar, JETP Lett. \textbf{97}, 249 (2013).

\bibitem{Yeh} P. Yeh, A. Yariv, and C.-S. Hong, J. Opt. Soc. Am. \textbf{67}, 423 (1977).

\bibitem {Morse-Feshbach} P.M. Morse and H. Feshbach. {\it Methods of theoretical physics. Part I.} (New York, Toronto, London, McGraw-Hill, 1953).

\bibitem{Quartz} R. Kitamura, L. Pilon, and M. Jonasz, Appl. Opt. \textbf{46}, 8118 (2007).

\bibitem{Teflon} P.J. van Zwol and G. Palasantzas, Phys. Rev. A. \textbf{81}, 062502 (2010).

\bibitem{Falk_JETP_2008} L.A. Falkovsky, JETP \textbf{106}, 575 (2008).

\bibitem{bilayer} M. Jablan, H. Buljan, and M. Solja\v{c}i\'{c}, Optics Express \textbf{19}, 11236 (2011).

\bibitem{Li} Z.Q. Li, E.A. Henriksen, Z. Jiang et al., Nat. Phys. \textbf{4}, 532 (2008).

\bibitem{Hwang} E.H. Hwang and S. Das Sarma, \prb \textbf{77}, 195412 (2008).

\end{references}
\end{document}